\documentclass[12pt,a4paper]{article}
\usepackage{graphicx}
\usepackage[T1]{fontenc}
\usepackage[utf8]{inputenc}
\usepackage{textcomp}
\usepackage[sc,osf]{mathpazo}
\usepackage{a4wide}  
\usepackage{latexsym,amsthm,amsfonts,amsmath,mathrsfs,amssymb}
\usepackage{dsfont}
\usepackage{accents}
\usepackage[nosort]{cite}
\usepackage{booktabs} 
\usepackage[unicode,implicit]{hyperref}
\hypersetup{%
  pdftitle    = {Komar integrals for theories of higher order in the curvature
  and black-hole chemistry}
  pdfkeywords = {Komar, Wald, conserved quantities, gravity, Noether, mass, entropy,  temperature, black hole, black-hole thermodynamics, black-hole chemistry},
  pdfauthor   = {Tom\'as Ort\'{\i}n},
  plainpages  = true,
  colorlinks  = true,
  citecolor   = blue,
  urlcolor    = red,
  linkcolor   = black
}
\newcommand{\hepth}[1]{{\tt
\href{http://www.arXiv.org/abs/hep-th/#1}{hep-th/#1}}}
\newcommand{\grqc}[1]{{\tt
\href{http://www.arXiv.org/abs/gr-qc/#1}{gr-qc/#1}}}

\newcommand{\arxiv}[1]{{\tt arXiv:\href{http://www.arXiv.org/abs/#1}{#1}}}
\newcommand{\doi}[1]{{\tt DOI:\href{http://dx.doi.org/#1}{#1}}}

\makeatletter
\@addtoreset{equation}{section}
\makeatother

\begin{document}

\begin{flushright}
\small
IFT-UAM/CSIC-21-42\\
April 21\textsuperscript{st}, 2021\\
\normalsize
\end{flushright}

\vspace{1cm}

\begin{center}

{\Large {\bf Komar integrals for theories of higher order\\[.5cm] in the
    Riemann curvature and black-hole chemistry}}

\vspace{1.5cm}

\renewcommand{\thefootnote}{\alph{footnote}}

{\sl\large  Tom\'{a}s Ort\'{\i}n}\footnote{Email: {\tt tomas.ortin[at]csic.es}}

\setcounter{footnote}{0}
\renewcommand{\thefootnote}{\arabic{footnote}}
\vspace{1.5cm}

{\it Instituto de F\'{\i}sica Te\'orica UAM/CSIC\\
C/ Nicol\'as Cabrera, 13--15,  C.U.~Cantoblanco, E-28049 Madrid, Spain}

\vspace{2.5cm}


{\bf Abstract}
\end{center}

\vspace{.5cm}

\begin{quotation}
  {\small We construct Komar-type integrals for theories of gravity of higher
    order in the Riemann curvature coupled to simple kinds of matter (scalar
    and vector fields) and we use them to compute Smarr formulae for
    black-hole solutions in those theories. The equivalence between $f(R)$ and
    Brans-Dicke theories is used to argue that the dimensionful parameters
    that appear in scalar potentials must be interpreted as thermodynamical
    variables (\textit{pressures}) and we give a general expression for their
    conjugate potentials (\textit{volumes}).
  }
\end{quotation}

\newpage
\pagestyle{plain}



\section*{Introduction}

Komar integrals \cite{Komar:1958wp} provide a simple and economic way of
computing the mass of spacetimes admitting a timelike Killing vector, but, as
shown in Ref.~\cite{Kastor:2010gq}, they can also be used to obtain
\textit{Smarr formulae} \cite{Smarr:1972kt} relating the mass to the conserved
charges. Since the Smarr formula can be obtains from the first law of
black-hole mechanics, the terms that occur in it have a direct thermodynamical
interpretation.\footnote{As shown in Ref.~\cite{Liberati:2015xcp}, Komar
  formulae can be obtained from the diffeomorphisms Noether charge using the
  methods of Refs.~\cite{Lee:1990nz,Wald:1993nt,Iyer:1994ys}.} This relation
and the presence of the cosmological constant in the Smarr formula
\cite{Kastor:2008xb,Kastor:2009wy} hinted at the interpretation of the
cosmological constant as a thermodynamical ``pressure'' with a conjugate
thermodynamical potential (``volume'') which, in its turn, means that the mass
should be interpreted as an enthalpy rather than as an internal energy. This
realization, extended to other dimensionful parameters occurring in the action
such as the coefficients of the Lovelock terms \cite{Kastor:2010gq}, has lead
to the discovery of a host of new phenomena involving black holes, opening a
new field that has been named \textit{black-hole chemistry}.\footnote{For
  reviews with many references see
  Refs.~\cite{Mann:2015luq,Kubiznak:2016qmn}.}

The coupling to matter has not been considered in most of these developments.
In particular, no Komar integrals have been proposed for theories with matter
fields, even though one simply has to follow the recipe of
Ref.~\cite{Liberati:2015xcp} to construct them systematically with the Noether
charge.\footnote{The recipe of Ref.~\cite{Liberati:2015xcp} leads to a
  combination of surface and volume integrals, but, as we will see, the volume
  integrals can always be rewritten as surface integrals.} One of the reasons
may be that the treatment of matter fields in
Refs.~\cite{Lee:1990nz,Wald:1993nt,Iyer:1994ys} is not valid for matter fields
that have some kind of gauge freedom (all matter fields but uncharged scalars,
as a matter of fact), which leads to Noether charges which are not
gauge-invariant, for instance.

In this paper, in Section~\ref{sec-Komarcharge}, we use the treatment of
fields with gauge freedoms proposed in
Refs.~\cite{Elgood:2020svt,Elgood:2020mdx,Elgood:2020nls} to find the Noether
charge and to construct explicit expressions for the Komar integral in
theories of gravity of higher order in the Riemann curvature minimally coupled
to a Maxwell field (see also \cite{Bueno:2016ypa}.) Additional fields of
different kinds can be treated in exactly the same way and we will work out in
Section~\ref{sec-examples} a few examples: General Relativity in presence of a
cosmological constant in Section~\ref{sec-GRL}, Lovelock gravities (studied in
\cite{Kastor:2010gq}) in Section~\ref{sec-Lovelock}, dilaton gravity in
Section~\ref{sec-dilatongravity} and $f(R)$ gravities in
Section~\ref{sec-f(R)gravity}.  In particular, and as a test of our formulae,
we will obtain the Smarr formula for cosmological
Reissner-Nordstr\"om-Tangherlini black holes (more often known as
$d$-dimensional Reissner-Nordstr\"om-(anti-)De Sitter black holes) in
Section~\ref{sec-CRNT}.

The example of $f(R)$ gravities is interesting and tractable. The action of
these theories contains a series of higher-order terms weighted by a
dimensionful parameter, as in general Lovelock gravities, which end up in the
Smarr formula. By analogy with the Lovelock case, it is natural to interpret
those parameters thermodynamical variables (``pressures'', again). The
equivalence between $f(R)$ and Brans-Dicke theories with a scalar potential
determined by the dimensionful parameters suggests that, in a general theory
with a scalar potential, the dimensionful parameters that define it must have
the same interpretation. It is, then, tempting to extend the analogy to all
the parameters that define the theory as a deformation of the simplest one
\cite{kn:EMMOP}. We will discuss this proposal and future work in
Section~\ref{sec-discussion}.

\section{Derivation of the Komar charge}
\label{sec-Komarcharge}

We are interested in general, diffeomorphism and gauge-invariant
$d$-dimensional, theories of metric gravity minimally coupled to arbitrary
matter fields which, in particular, can have gauge freedoms. However, for the
sake of simplicity, we will just consider a Maxwell field $A_{\mu}$ with field
strength $F_{\mu\nu}=2\partial_{[\mu}A_{\nu]}$, which suffices to illustrate
the treatment of the gauge transformations. We also set the overall factor of
$16\pi G_{N}^{(d)}=1$ for the moment. Thus, we consider the action

\begin{equation}
  S[g,A]
  =
  \int d^{d}x\left\{\mathcal{L}_{\rm grav}-\tfrac{1}{4}\sqrt{|g|}F^{2}\right\}\,.
\end{equation}

\noindent
We assume that $S_{\rm grav}[g]$ contains terms of arbitrary order in the
Riemann curvature tensor contracted using the metric.\footnote{In our
  conventions, the Riemann tensor is defined as
\begin{equation}
  R_{\mu\nu\rho}{}^{\sigma}
  \equiv
  2\partial_{[\mu}\Gamma_{\nu]\rho}{}^{\sigma}
  +2\Gamma_{[\mu|\lambda}{}^{\sigma}\Gamma_{|\nu]\rho}{}^{\lambda}\,,
\end{equation}
where $\Gamma_{\mu\nu}{}^{\rho}$ is the Levi-Civita connection, whose
components are the Christoffel symbols
\begin{equation}
  \Gamma_{\mu\nu}{}^{\rho}
  =
  \biggl\{\!\!
\begin{array}{c}
\rho\\
\mu\, \nu \\
\end{array}
\!\!
\biggr\}
\equiv
\tfrac{1}{2}g^{\rho\sigma}\!\!
\left\{\partial_{\mu}g_{\nu\sigma} +\partial_{\nu}g_{\mu\sigma}
-\partial_{\sigma}g_{\mu\nu}\right\}\,.
\end{equation}
} 

Under an arbitrary variation of the fields, the action behaves as

\begin{equation}
  \label{eq:generalvariation}
  \delta S
  =
  \int \left\{\frac{\delta S}{\delta g_{\mu\nu}}\delta g_{\mu\nu}
    +\frac{\delta S}{\delta A_{\mu}}\delta A_{\mu}
    +\partial_{\mu}\Theta^{\mu}(g,\delta_{\xi}g,A,\delta_{\xi}A)\right\}\,.
\end{equation}

\noindent
Using the formalism of Ref.~\cite{Padmanabhan:2011ex},\footnote{See also
  Ref.~\cite{Bueno:2016ypa} for many results related to higher-order theories,
  in particular in relatio nto Wald's formalism.} the equations of motion and
the total derivative can be written in the form

\begin{subequations}
  \begin{align}
    \label{eq:EinsteinEquations}
    \frac{\delta S}{\delta g_{\mu\nu}}
    & =
      \tfrac{1}{2}g^{\mu\nu}\mathcal{L}_{\rm grav}
      -P^{\mu\alpha\beta\gamma}R^{\nu}{}_{\alpha\beta\gamma}
      +2\nabla_{(\alpha}\nabla_{\beta)}P^{\alpha\mu\beta\nu}
      +\tfrac{1}{2}\sqrt{|g|}\, T^{\mu\nu}\,,
    \\
    & \nonumber \\
    \frac{\delta S}{\delta A_{\mu}}
    & =
      \partial_{\mu}\left(\sqrt{|g|}F^{\mu\nu}\right)\,,      
    \\
    & \nonumber \\
\Theta^{\mu}(g,\delta g,A,\delta A)
    & =
      2P^{\mu\alpha\beta}{}_{\gamma}\delta \Gamma_{\alpha\beta}{}^{\gamma}
      -2\nabla_{\gamma}P^{\gamma\alpha\mu\beta}\delta g_{\alpha\beta}
      -\sqrt{|g|}F^{\mu\nu}\delta A_{\nu}\,.
  \end{align}
\end{subequations}

\noindent
with

\begin{subequations}
  \begin{align}
    P^{\mu\nu\rho\sigma}
    & \equiv
      \frac{\partial \mathcal{L}_{\rm grav}}{\partial R_{\mu\nu\rho\sigma}}\,,
    \\
    & \nonumber \\
    T^{\mu\nu}
  & \equiv
F^{\mu\alpha}F^{\nu}{}_{\alpha}-\tfrac{1}{4}g^{\mu\nu}F^{2}\,,
    \\
    & \nonumber \\
\delta \Gamma_{\alpha\beta}{}^{\gamma}    
    & =
      \tfrac{1}{2}g^{\gamma\delta}\{\nabla_{\alpha}\delta g_{\beta\delta}
      +\nabla_{\beta}\delta g_{\alpha\delta}-\nabla_{\delta}\delta g_{\alpha\beta}\}\,.
  \end{align}
\end{subequations}

\noindent
It follows from its definition that $P^{\mu\nu\rho\sigma}$ shares the
symmetries of the Riemann tensor which, in particular, means that

\begin{equation}
  \label{eq:BianchiP}
P^{[\mu\nu\rho]\sigma}=0\,.  
\end{equation}

Another important property is

\begin{equation}
  \label{eq:PRproperty}
P^{[\mu|\alpha\beta\gamma}R^{|\nu]}{}_{\alpha\beta\gamma}=0\,.
\end{equation}

Now, we specialize this general expression to the case of diffeomorphisms
generated by an infinitesimal vector field $\xi^{\mu}$. The transformations of
the fields are given by

\begin{subequations}
  \begin{align}
    \delta_{\xi}g_{\mu\nu}
    & =
      -\pounds_{\xi}g_{\mu\nu}
      =
      -2\nabla_{(\mu}\xi_{\nu)}\,,
    \\
    & \nonumber \\
    \delta_{\xi}A_{\mu}
    & =
      -\mathbb{L}_{\xi}A_{\mu}\,,
  \end{align}
\end{subequations}

\noindent
where $\mathbb{L}_{\xi}$ is a generalization of the standard Lie derivative
$\pounds_{\xi}$ that takes into account the gauge freedoms of the matter
fields and which is a combination of the standard Lie derivative
$\pounds_{\xi}$ and ``compensating'' gauge transformations with
$\xi$-dependent gauge parameters $\chi_{\xi}$, $\delta_{\chi_{\xi}}$.\footnote{See
  Ref.~\cite{Elgood:2020svt} and references therein.}  In the case of a
Maxwell field,

\begin{equation}
\chi_{\xi}= \xi^{\mu}A_{\mu}-P_{\xi}\,,
\end{equation}

\noindent
where $P_{\xi}$ is such that, when $\xi$ generates a symmetry of the whole
field configuration (in particular, it is a Killing vector), it satisfies the
momentum-map equation

\begin{equation}
  \label{eq:momentummap}
\partial_{\mu} P_{k} = - k^{\nu}F_{\mu\nu}\,. 
\end{equation}

Then,

\begin{equation}
  \begin{aligned}
    \mathbb{L}_{\xi}A_{\mu}
    & =
    \pounds_{\xi}A_{\mu}-\delta_{\chi_{\xi}}A_{\mu}
    = \xi^{\nu}\partial_{\nu}A_{\mu}+\partial_{\mu}\xi^{\nu}A_{\nu}
    -\partial_{\mu}(\xi^{\nu}A_{\nu}-P_{\xi})
    \\
    & \\
    & =
    \xi^{\nu}F_{\nu\mu}+\partial_{\mu}P_{\xi}\,,
  \end{aligned}
\end{equation}

\noindent
which vanishes identically for $\xi^{\mu}=k^{\mu}$ by virtue of the
momentum-map Eq.~(\ref{eq:momentummap}).

Substituting these variations in Eq.~(\ref{eq:generalvariation}) we get

\begin{equation}
  \label{eq:xivariation0}
    \delta_{\xi} S
     =
    -\int \left\{2\frac{\delta S}{\delta
        g_{\mu\nu}}\nabla_{\mu}\xi_{\nu} +\frac{\delta S}{\delta
        A_{\mu}}\left(\xi^{\nu}F_{\nu\mu}+\partial_{\mu}P_{\xi}\right)
      -\partial_{\mu}\Theta^{\mu}(g,\delta_{\xi}g,A,\delta_{\xi}A)\right\}\,,
\end{equation}

\noindent
where

\begin{equation}
  \label{eq:Thetaxi}
  \begin{aligned}
    \Theta^{\mu}(g,\delta_{\xi}g,A,\delta_{\xi}A)
    & \equiv
    2P^{\mu\alpha\beta}{}_{\gamma}\delta \Gamma_{\alpha\beta}{}^{\gamma}
    -2\nabla_{\gamma}P^{\gamma\alpha\mu\beta}\delta g_{\alpha\beta}
    -\sqrt{|g|}F^{\mu\nu}\delta A_{\nu}
    \\
    & \\
    & =
    -4P^{\mu\alpha\beta\gamma}\nabla_{\beta}\nabla_{(\alpha}\xi_{\gamma)}
    +4\nabla_{\gamma}P^{\gamma\alpha\mu\beta}\nabla_{(\alpha}\xi_{\beta)}
    \\
    & \\
    & \hspace{.5cm}
    +\sqrt{|g|}F^{\mu\alpha}F_{\nu\alpha}\xi^{\nu}
    +\sqrt{|g|}F^{\mu\nu}\partial_{\nu}P_{\xi}\,.
  \end{aligned}
\end{equation}

Integrating by parts the first term and using the identity

\begin{equation}
  \nabla_{\mu}\frac{\delta S}{\delta g_{\mu\nu}}
  =
  \tfrac{1}{2}\sqrt{|g|}\nabla_{\mu}T^{\mu\nu}
  =
  \tfrac{1}{2}\frac{\delta S}{\delta A_{\alpha}}F_{\nu\alpha}\,,
\end{equation}

\noindent
we get

\begin{equation}
  \label{eq:xivariation1}
    \delta_{\xi} S
     =
     -\int \left\{
\frac{\delta S}{\delta A_{\mu}}\partial_{\mu}P_{\xi}
-\partial_{\mu}\left[\Theta^{\mu}(g,\delta_{\xi}g,A,\delta_{\xi}A)
        -2\frac{\delta S}{\delta g_{\mu\nu}}\xi^{\nu}\right]\right\}\,.
\end{equation}

Integrating by parts again and using now the Noether identity associated to
the invariance under the gauge transformations $\delta_{\chi}A_{\mu} =
\partial_{\mu}\chi$, namely

\begin{equation}
\partial_{\mu}\frac{\delta S}{\delta A_{\mu}}=0\,,
\end{equation}

\noindent
we arrive at

\begin{equation}
  \label{eq:xivariation2}
    \delta_{\xi} S
     =
     \int 
\partial_{\mu}\left[\Theta^{\mu}(g,\delta_{\xi}g,A,\delta_{\xi}A)
        -2\frac{\delta S}{\delta g_{\mu\nu}}\xi_{\nu}
      -\frac{\delta S}{\delta A_{\mu}}P_{\xi}\right]\,.
\end{equation}

Finally, the invariance of the action under diffeomorphisms means that

\begin{equation}
\delta_{\xi} S = -\int d^{d}x \partial_{\mu}\left(\xi^{\mu}\mathcal{L}\right)\,,
\end{equation}

\noindent
and we arrive at the following off-shell identity:

\begin{equation}
  \label{eq:thetaprime}
\partial_{\mu} J^{\mu}=0\,,
\end{equation}

\noindent
where

\begin{equation}
  \label{eq:Jmu}
  \begin{aligned}
 J^{\mu}
 & =
\Theta^{\mu}(g,\delta_{\xi}g,A,\delta_{\xi}A)
        -2\frac{\delta S}{\delta g_{\mu\nu}}\xi_{\nu}
        -\frac{\delta S}{\delta A_{\mu}}P_{\xi}
        +\xi^{\mu}\mathcal{L}
        \\
        & \\
        & =
    -4P^{\mu\alpha\beta\gamma}\nabla_{\beta}\nabla_{(\alpha}\xi_{\gamma)}
    +4\nabla_{\gamma}P^{\gamma\alpha\mu\beta}\nabla_{(\alpha}\xi_{\beta)}
      +2P^{\mu\alpha\beta\gamma}R^{\nu}{}_{\alpha\beta\gamma}\xi_{\nu}
      -4\nabla_{(\alpha}\nabla_{\beta)}P^{\alpha\mu\beta\nu}\xi_{\nu}
    \\
    & \\
    & \hspace{.5cm}
    -\partial_{\nu}\left(\sqrt{|g|}F^{\nu\mu}P_{\xi}\right)\,.
          \end{aligned}
\end{equation}

As expected, locally, one can find an antisymmetric $J^{\alpha\mu}$ such
that\footnote{The detailed calculation can be found in the Appendix.}

\begin{equation}
J^{\mu}=\partial_{\alpha}J^{\alpha\mu}\,,
\end{equation}

\noindent
and, restoring the overall normalization factor of $(16\pi G_{N}^{(d)})^{-1}$
in the action, it is given by

\begin{equation}
  \label{eq:Jalphamu}
  J^{\alpha\mu}
  =
  -4\nabla_{\beta}P^{\alpha\mu\beta\nu}\xi_{\nu}
  +2P^{\alpha\mu\beta\nu}\nabla_{\beta}\xi_{\nu}
  -\frac{1}{16\pi G_{N}^{(d)}}\sqrt{|g|}F^{\alpha\mu}P_{\xi}\,.
\end{equation}

Now, for a Killing vector that generates an isometry which also leaves
invariant the Maxwell field so we can use the momentum-map equation
(\ref{eq:momentummap})

\begin{equation}
  \begin{aligned}
    J^{\mu}
    &
    =
    \left(2P^{\mu\alpha\beta\gamma}R^{\nu}{}_{\alpha\beta\gamma}
    -4\nabla_{\alpha}\nabla_{\beta}P^{\alpha\mu\beta\nu}\right)k_{\nu}
    -\frac{1}{16\pi G_{N}^{(d)}}\partial_{\nu}\left(\sqrt{|g|}F^{\nu\mu}P_{k}\right)
    \\
    & \\
    &
    =
      \mathcal{L}k^{\mu}\,.
  \end{aligned}
\end{equation}

\noindent
where we have used the equations of motion and the momentum-map equation

Generically, this current does not vanish on-shell because the Lagrangian is
not guaranteed to vanish on-shell. The fact that it does in pure Einstein
gravity is what makes it so easy to find the Komar integral in that case.

In the case of general Lovelock theories, where this problem arises even in
absence of matter coupling one may try to include volume terms to the Komar
surface integral, as proposed in Ref.~\cite{Kastor:2008xb}. There is, however,
another possibility inspired in the solution given in the same reference to
the cosmological constant term: we can modify $J^{\alpha\mu}$ to absorb that
term

\begin{equation}
  \tilde{J}^{\alpha\mu}
  =
  J^{\alpha\mu} -2\omega^{\alpha\mu}\,,
\end{equation}

\noindent
where $\omega^{\alpha\mu}$ is a generalization of the Killing potential
(density) introduced in Ref.~\cite{Kastor:2008xb} defined by

\begin{equation}
\partial_{\alpha}\omega^{\alpha\mu} =\tfrac{1}{2}\mathcal{L}k^{\mu}\,,   
\end{equation}

\noindent
and whose local existence is guaranteed by the Killing equation and the
symmetry condition

\begin{equation}
k^{\mu}\partial_{\mu}\mathcal{L}=0\,,
\end{equation}

\noindent
which must be satisfied if the diffeomorphism generated by $k^{\mu}$ is a
symmetry of the complete solution.

We arrive, then, to the generalization of the Komar integral we were looking for:

\begin{equation}
  \label{eq:Komarintegralgeneral}
  \begin{aligned}
    \mathcal{K}(\Sigma^{d-2})
    & \equiv
    \tfrac{(-1)^{d-1}}{2}\int_{\Sigma^{d-2}} \frac{d^{d-2}\Sigma_{\alpha\mu}}{\sqrt{|g|}}\tilde{J}^{\alpha\mu}
    \\
    & \\
    & =  (-1)^{d-1}\int_{\Sigma^{d-2}}\frac{d^{d-2}\Sigma_{\alpha\mu}}{\sqrt{|g|}} \left\{
      P^{\alpha\mu\beta\nu}\nabla_{\beta}k_{\nu}
      -2\nabla_{\beta}P^{\alpha\mu\beta\nu}k_{\nu}
      -\frac{1}{32\pi G_{N}^{(d)}}\sqrt{|g|}F^{\alpha\mu}P_{k}
      -\omega^{\alpha\mu} \right\}\,.
  \end{aligned}
\end{equation}

Although we have only considered one matter field, it is not difficult to
adapt this formula to include an uncharged scalar field with a scalar
potential or a cosmological constant term, which will only contribute to the
Killing potential. In the Section~\ref{sec-examples} we are going to study a
few examples.

\section{Examples}
\label{sec-examples}

\subsection{General Relativity in presence of a cosmological constant}
\label{sec-GRL}

The action of this theory is

\begin{equation}
  S[g]
  =
  \frac{1}{16\pi G_{N}^{(d)}}
  \int d^{d}x\sqrt{|g|}\left\{R -(d-2)\Lambda\right\}\,.
\end{equation}

\noindent
We have normalized the cosmological constant so that the equation of motion
is\footnote{Our cosmological constant is, therefore, related to that in
  Ref.~\cite{Kastor:2009wy} by $\Lambda_{KRT}=\frac{(d-2)}{2}\Lambda$.}

\begin{equation}
R_{\mu\nu}=\Lambda g_{\mu\nu}\,,  
\end{equation}

\noindent
in any dimension. This means that, on-shell,

\begin{equation}
\mathcal{L}/\sqrt{|g|}= 2\Lambda\,.
\end{equation}

\noindent
We can define $\omega^{\alpha\mu}$ by the equation

\begin{equation}
  \nabla_{\alpha}\omega^{\alpha\mu}
  =
  \Lambda k^{\mu}\,,
\end{equation}

\noindent
to get a simpler expression.

The $P^{\alpha\mu\beta\nu}$  associated to the Einstein-Hilbert term is

\begin{equation}
  \label{eq:EHP}
  P^{\alpha\mu\beta\nu}
  =
  \frac{1}{16\pi G_{N}^{(d)}}\sqrt{|g|}g^{\alpha\mu\, \beta\nu}\,,
  \,\,\,\,\,\,
\text{where}
  \,\,\,\,\,\,
  g^{\alpha\mu\, \beta\nu}
  \equiv
  \tfrac{1}{2}\left(g^{\alpha\beta}g^{\mu\nu}-g^{\alpha\nu}g^{\mu\beta}\right)\,,
\end{equation}

\noindent
and taking into account all these terms we arrive to the integral

\begin{equation}
\mathcal{K}(\Sigma^{d-2})
    =
    \frac{(-1)^{d-1}}{16\pi G_{N}^{(d)}}\int_{\Sigma^{d-2}} d^{d-2}\Sigma_{\alpha\mu} 
    \left[ \nabla^{\alpha}\xi^{\mu}-\omega^{\alpha\mu}\right]\,,
\end{equation}

\noindent
which, up to normalization, is the integral proposed in
Ref.~\cite{Kastor:2008xb} and which reduces to the standard Komar integral
\cite{Komar:1958wp} in absence of cosmological constant.

We can use this Komar integral to find the Smarr formula for
Schwarzschild-(a-)DS black holes following
Ref.~\cite{Kastor:2010gq,Liberati:2015xcp}, but we will obtain it as a
particular case of the Smarr formula for cosmological
Reissner-Nordstr\"om-Tangherlini black holes that we will derive in
Section~\ref{sec-CRNT}.

\subsection{Lovelock gravities}
\label{sec-Lovelock}

Lovelock gravities are characterized by the Lagrangian densities
\cite{Lovelock:1971yv} 

\begin{equation}
  \begin{aligned}
    \mathcal{L}^{k}
    & =
    \frac{\sqrt{|g|}}{16\pi G_{N}{}^{(d)}}
    g^{\alpha_{1}\cdots\alpha_{2k}\, \beta_{1}\cdots\beta_{2k}}
    R_{\alpha_{1}\alpha_{2}\, \beta_{1}\beta_{2}}\cdots
    R_{\alpha_{2k-1}\alpha_{2k}\, \beta_{2k-1}\beta_{2k}}
    \\
    & \\
    & =
    \frac{\sqrt{|g|}}{16\pi G_{N}{}^{(d)}}
    R_{\alpha_{1}\alpha_{2}}{}^{[\alpha_{1}\alpha_{2}}\cdots
    R_{\alpha_{2k-1}\alpha_{2k}}{}^{\alpha_{2k-1}\alpha_{2k}]}\,,
  \end{aligned}
\end{equation}

\noindent
where $k=0,1,\ldots$, which are non-trivial kinetic terms for the metric only
for $0<2k<d$. The case $k=0$ corresponds to the cosmological constant term of
the previous example with $\Lambda=-1/(d-2)$. It plays a non-trivial role when
it is combined with other terms. 

We just need to compute

\begin{subequations}
  \begin{align}
  P_{k}^{\mu\nu\rho\sigma}
  & =
  \frac{k\sqrt{|g|}}{16\pi G_{N}{}^{(d)}}
  g^{\mu\nu\alpha_{3}\cdots\alpha_{2k}\, \rho\sigma\beta_{3}\cdots\beta_{2k}}
  R_{\alpha_{3}\alpha_{4}\, \beta_{3}\beta_{4}}\cdots
    R_{\alpha_{2k-1}\alpha_{2k}\, \beta_{2k-1}\beta_{2k}}
    \nonumber \\
    & \nonumber \\
    & =
        \frac{k\sqrt{|g|}}{16\pi G_{N}{}^{(d)}}
  g^{\mu\nu\, [\rho\sigma}
  R_{\alpha_{3}\alpha_{4}}{}^{\alpha_{3}\alpha_{4}}\cdots
    R_{\alpha_{2k-1}\alpha_{2k}}{}^{\alpha_{2k-1}\alpha_{2k}]}\,,
    \\
    & \nonumber \\ 
\nabla_{\mu}  P^{\mu\nu\rho\sigma}
  & =
0\,,
  \end{align}
\end{subequations}

\noindent
to get the equations of motion (up to an overall factor)

\begin{equation}
  k\mathcal{G}^{k\, \mu\nu}-\tfrac{1}{2}g^{\mu\nu}\mathcal{G}^{k\, \rho}{}_{\rho}
  =
  0\,,
\end{equation}

\noindent
where

\begin{subequations}
  \begin{align}
    \mathcal{G}^{k\, \mu\nu}
    & \equiv
      g^{\mu\rho\, [\sigma\tau}
    R_{\alpha_{3}\alpha_{4}}{}^{\alpha_{3}\alpha_{4}}\cdots
      R_{\alpha_{2k-1}\alpha_{2k}}{}^{\alpha_{2k-1}\alpha_{2k}]}R^{\nu}{}_{\rho\sigma\tau}\,,
    \\
    & \nonumber \\
    \Rightarrow\,\,\,\, \mathcal{G}^{k\, \rho}{}_{\rho}
    & =
    \frac{16\pi G_{N}{}^{(d)}}{\sqrt{|g|}}\mathcal{L}^{k}\,.
  \end{align}
\end{subequations}

The trace is proportional to the Lagrangian and, therefore, the equations
of motion are equivalent to 

\begin{equation}
 \mathcal{G}^{k\, \mu\nu}
  =
  0\,,
\end{equation}

\noindent
which means that the Lagrangian vanishes on-shell. This leads immediately to

\begin{equation}
  \begin{aligned}
    \mathcal{K}^{k}(\Sigma^{d-2})
    & =
    \frac{(-1)^{d-1}k}{16\pi G_{N}{}^{(d)}}\int_{\Sigma^{d-2}}
    d^{d-2}\Sigma_{\alpha_{1}\alpha_{2}} \nabla^{[\alpha_{1}}\xi^{\alpha_{2}}    
  R_{\alpha_{3}\alpha_{4}}{}^{\alpha_{3}\alpha_{4}}\cdots
    R_{\alpha_{2k-1}\alpha_{2k}}{}^{\alpha_{2k-1}\alpha_{2k}]}\,,
  \end{aligned}
\end{equation}

\noindent
which is, up to normalization, the Komar integral proposed in
Ref.~\cite{Kastor:2008xb}.

If we consider a linear combination of Lovelock terms with constant
coefficients $\alpha_{k}$\footnote{Now, as we said, the $k=0$ term becomes
  relevant and the constant $\alpha_{0}=-(d-2)\Lambda$.}

\begin{equation}
\mathcal{L}=\sum_{k=0}\alpha_{k}\mathcal{L}^{k}\,,  
\end{equation}

\noindent
the equations of motion are linear combinations of the $k$th equations of
motion

\begin{equation}
  \sum_{k}\alpha_{k}
  \left[\mathcal{G}^{k\, \mu\nu}-\tfrac{1}{2k}g^{\mu\nu}\mathcal{G}^{k\,
      \rho}{}_{\rho} \right]
  =
  0\,,
\end{equation}

\noindent
and the trace, which is not proportional to the Lagrangian anymore, gives the
equation

\begin{equation}
\mathcal{L}
  \stackrel{\rm on-shell}{=}
  \frac{2}{d}  \sum_{k=0}k\alpha_{k} \mathcal{L}^{k}\,.
\end{equation}

This relation can be used to eliminate one term from $\mathcal{L}$, typically
the one of highest order, $k=m$. If $m\neq d/2$, the result is

\begin{equation}
\mathcal{L}
  \stackrel{\rm on-shell}{=}
  \frac{2}{d}  \sum^{m\neq d/2}_{k}\frac{2(m-k)}{2m-d}\alpha_{k} \mathcal{L}^{k}\,.
\end{equation}

\noindent
If $m= d/2$, the highest order term is topological and we can eliminate the
next one, of order $k=m-1$. The result is

\begin{equation}
\mathcal{L}
  \stackrel{\rm on-shell}{=}
  -\sum^{d/2}_{k}2(d-2-2k)\alpha_{k} \mathcal{L}^{k}\,.
\end{equation}

The Komar integral for these theories takes, then, the form

\begin{equation}
  \mathcal{K}(\Sigma^{d-2})
  =
\left\{
  \begin{array}{l}
{\displaystyle\sum^{m\neq d/2}_{k}\alpha_{k} \mathcal{K}^{k}(\Sigma^{d-2})
  +(-1)^{d}\sum^{m\neq d/2}_{k}\frac{2(m-k)}{2m-d}\alpha_{k}
    \int_{\Sigma^{d-2}}d^{d-2}\Sigma_{\alpha\mu}\omega^{k\,\alpha\mu}}\,,
    \\
    \\
{\displaystyle\sum^{m=d/2}_{k}\alpha_{k} \mathcal{K}^{k}(\Sigma^{d-2})
  +(-1)^{d-1}\sum^{d/2}_{k}2(d-2-2k)\alpha_{k}
    \int_{\Sigma^{d-2}}d^{d-2}\Sigma_{\alpha\mu}\omega^{k\,\alpha\mu}}\,,    
  \end{array}
  \right.
\end{equation}

\noindent
where the $\omega^{k\,\alpha\mu}$ satisfy the equations

\begin{equation}
  \nabla_{\alpha}\omega^{k\,\alpha\mu}
  =
  \frac{16\pi G_{N}{}^{(d)}}{\sqrt{|g|}}\mathcal{L}^{k}\,,
\end{equation}

\noindent
where each of the Lagrangian densities $\mathcal{L}^{k}$ has to be evaluated
on-shell.

Using these Komar integrals we can recover the results of
Refs.~\cite{Kastor:2010gq,Liberati:2015xcp} on the Smarr formula for black
holes in these theories. We refer the reader to those articles for further
details.

\subsection{Dilaton gravity}
\label{sec-dilatongravity}

The action of dilaton gravity, to which we have added a scalar potential, is
given (in the Einstein frame) by 

\begin{equation}
  S[g,A,\phi]
  =
  \frac{1}{16\pi G_{N}^{(d)}}
  \int d^{d}x\sqrt{|g|}\left\{R
    +2(\partial\phi)^{2}-\tfrac{1}{4}e^{-2a\phi}F^{2} -V(\phi)\right\}\,,
\end{equation}

\noindent
where $a$ is a constant parameter. This example includes the one considered in
Section~\ref{sec-GRL} if we set to zero the scalar and Maxwell fields and
replace the scalar potential by $(d-2)\Lambda$.

$P^{\mu\nu\rho\sigma}$ is, again, the one
associated to the Einstein-Hilbert term Eq.~(\ref{eq:EHP}). The scalar field
has no gauge freedom and, therefore,

\begin{equation}
  \label{eq:Komarintegraldilatongravity}
\mathcal{K}(\Sigma^{d-2})
    =
    \frac{(-1)^{d-1}}{16\pi G_{N}^{(d)}}\int_{\Sigma^{d-2}} d^{d-2}\Sigma_{\alpha\mu} 
    \left[ \nabla^{\alpha}\xi^{\mu}-\omega^{\alpha\mu}
      -\tfrac{1}{2}e^{-2a\phi}F^{\alpha\mu}P_{k}\right]\,,
\end{equation}

\noindent
where $P_{k}$ satisfies the same momentum-map equation (\ref{eq:momentummap})
but, now, $\omega^{\alpha\mu}$ satisfies the equation

\begin{equation}
  \nabla_{\alpha}\omega^{\alpha\mu}
  =
  \frac{8\pi G_{N}^{(d)}}{\sqrt{|g|}}\mathcal{L} k^{\mu}\,,   
\end{equation}

\noindent
where the Lagrangian 

\begin{equation}
  \frac{8\pi G_{N}^{(d)}}{\sqrt{|g|}}\mathcal{L}
  =
  \tfrac{1}{2}\left\{R
    +2(\partial\phi)^{2}-\tfrac{1}{4}e^{-2a\phi}F^{2} -V(\phi)\right\}\,,
\end{equation}

\noindent
has to be evaluated on-shell.



The trace of the Einstein equation gives

\begin{equation}
  R+2(\partial\phi)^{2}=
    \frac{(d-4)}{4(d-2)}e^{-2a\phi}F^{2}
    +\frac{d}{(d-2)}V(\phi)\,,
\end{equation}

\noindent
and the on-shell Lagrangian is given by

\begin{equation}
  \frac{8\pi G_{N}^{(d)}}{\sqrt{|g|}}\mathcal{L}
  \stackrel{on-shell}{=}
  -\frac{1}{2(d-2)}e^{-2a\phi}F^{2}
    +\frac{1}{(d-2)}V(\phi)\,.
\end{equation}

Non-trivial asymptotically-De Sitter and anti-De Sitter dilaton black holes
with an \textit{ad hoc} dilaton potential in 4 and higher dimensions have been
constructed in Refs.~\cite{Gao:2004tu} and \cite{Gao:2004tv},
respectively. They are quite complicated to handle and, therefore, we will
just consider the $d$-dimensional cosmological
Reissner-Nordstr\"om-Tangherlini black holes (more often known as
$d$-dimensional Reissner-Nordstr\"om-(anti-)De Sitter black holes)
\cite{Tangherlini:1963bw}. In Section~\ref{sec-f(R)gravity} we will consider
also Schwarzschild-(a-)DS solutions.

\subsubsection{Cosmological Reissner-Nordstr\"om-Tangherlini black holes}
\label{sec-CRNT}

These solutions take the simple form

\begin{equation}
  \label{eq:CRNTsolution}
\begin{aligned}
  ds^{2}
  & =
  Wdt^{2} -W^{-1}dr^{2} -r^{2}d\Omega_{(d-2)}^{2}\,,
  \\
  & \\
  W
  & =
  1 -\frac{2m}{r^{d-3}}+\left(\frac{q}{r^{d-3}}\right)^{2}
  -\frac{\Lambda}{(d-1)}r^{2}\,,
  \\
  & \\
  A
  & =
  \sqrt{\frac{2(d-2)}{(d-3)}}
  \frac{q}{r^{d-3}}dt\,,
\end{aligned}
\end{equation}

\noindent
where the parameters $m$ and $q$, introduced to simplify the expressions, are
related to the ADM mass, $M$ and to the canonically-normalized electric
charge, $Q$, by

\begin{equation}
  \label{eq:CRNTphysicalparameters}
  m = \gamma M\,,
  \hspace{1cm}
  q= \sqrt{\frac{2(d-2)}{(d-3)}} \gamma  Q\,,
  \hspace{1cm}
  \gamma \equiv \frac{8\pi G_{N}^{(d)}}{(d-2)\omega_{(d-2)}}\,.
\end{equation}

We are going to focus on the aDS ($\Lambda <0$) case for simplicity, since in
that case there is only one horizon at $r=r_{h}$, which is the value of the
$r$ at which $W(r_{h})=0$.

In order to find the Smarr formula for these black holes, following
Ref.~\cite{Kastor:2010gq,Liberati:2015xcp}, we integrate the divergence of the
integrand of the Komar integral (which vanishes identically, by construction)
on a hypersurface whose boundary is the disjoint union of a spatial section of
the event horizon and spatial infinity. Stokes' theorem tells us that the
Komar integrals over the two disjoint components of the boundary must be
equal. This is the basis of the Smarr formula. Let us compute each of these
integrals.

First of all, the Komar integral can be obtained from
Eq.~(\ref{eq:Komarintegraldilatongravity}) and reads

\begin{equation}
  \label{eq:KomarintegralcomoEM}
\mathcal{K}(\Sigma^{d-2})
    =
    \frac{(-1)^{d-1}}{8\pi G_{N}^{(d)}}\int_{\Sigma^{d-2}} d^{d-2}\Sigma_{\alpha\mu} 
    \left[ \nabla^{\alpha}\xi^{\mu}-\omega^{\alpha\mu}
      -\tfrac{1}{4}F^{\alpha\mu}P_{k}\right]\,,
\end{equation}

\noindent
where $\omega^{\alpha\mu}$ satisfies

\begin{equation}
  \nabla_{\alpha}\omega^{\alpha\mu}
  =
\left[ -\frac{1}{4(d-2)}F^{2}
    +\Lambda\right]k^{\mu}\,,  
\end{equation}

\noindent
and where $k^{\mu}=\delta^{\mu}{}_{t}$. This equation reduces to

\begin{equation}
  \frac{1}{r^{d-2}}\left(r^{d-2}\omega^{rt}\right)'
  =
  \frac{(d-3)q^{2}}{r^{2(d-2)}}+\Lambda\,,
\end{equation}

\noindent
and is solved by

\begin{equation}
  \begin{aligned}
    -\omega^{tr}
    & =
    -\frac{q^{2}}{r^{2(d-3)+1}}+\frac{\Lambda}{(d-1)} r +\alpha
    r_{h}\left(\frac{r_{h}}{r}\right)^{d-2}\,,
  \end{aligned}
\end{equation}

\noindent
where the last term has been added, using the notation of
Ref.~\cite{Kastor:2008xb}, to reflect the possibility of adding the dual of an
exact $(d-2)$-form to $\omega$.

The momentum map equation (\ref{eq:momentummap}) is solved by $P_{k}=A_{t}$
and

\begin{equation}
  -\tfrac{1}{2}F^{tr}P_{k}
  =
  \frac{(d-2)q^{2}}{r^{2(d-3)+1}}\,.
\end{equation}

Finally,

\begin{equation}
  \nabla^{[t}k^{r]}
  =
  \tfrac{1}{2}W' =
  \frac{(d-3)m}{r^{d-2}} -\frac{(d-2)q^{2}}{2r^{2(d-3)+1}} -\frac{\Lambda}{(d-1)}r\,,
\end{equation}

\noindent
and, integrating over a sphere of constant radius $r$

\begin{equation}
(-1)^{d-1}\int_{S_{r}^{d-2}} d^{d-2}\Sigma_{tr}
  =
  \omega_{(d-2)}r^{d-2}\,,
\end{equation}

\noindent
we have

\begin{equation}
  \begin{aligned}
    \mathcal{K}(S_{r}^{d-2})
    & =
    \frac{(-1)^{d-1}}{8\pi
      G_{N}^{(d)}}\int_{S_{r}^{d-2}} d^{d-2}\Sigma_{tr}
    \left[\frac{(d-3)m}{r^{d-2}}
      +\frac{(d-4)q^{2}}{2(d-3)r^{2(d-3)+1}}
      +\alpha r_{h}\left(\frac{r_{h}}{r}\right)^{d-2}\right]
    \\
    & \\
    & =
    \left[\frac{(d-3)M}{(d-2)}
      +\frac{(d-4)\gamma Q^{2}}{(d-3)^{2}r^{(d-3)}}
      +\frac{\alpha r_{h}^{d-1}}{(d-2)\gamma }\right]
    \stackrel{r\rightarrow \infty}{\longrightarrow}
    \frac{(d-3)M}{(d-2)}+\frac{\alpha r_{h}^{d-1}}{(d-2)\gamma }\,.
  \end{aligned}
\end{equation}

The integral on the horizon at $r=r_{h}$ is easy to compute realizing that

\begin{equation}
  \tfrac{1}{2}W'(r_{h}) = \kappa\,,
\end{equation}

\noindent
the surface gravity. Then, 

\begin{equation}
  \begin{aligned}
    \mathcal{K}(S_{r_{h}}^{d-2})
    & =
    \frac{\omega_{(d-2)}r_{h}^{d-2}}{8\pi
      G_{N}^{(d)}}
    \left[\kappa 
      +\frac{(d-3)q^{2}}{r_{h}^{2(d-3)+1}}+\frac{\Lambda}{(d-1)} r_{h}
      +\alpha r_{h}
      \right]
    & =
    \frac{ \kappa A}{8\pi G_{N}^{(d)}}
    +\frac{(d-3)}{(d-2)}\Phi Q
    -2\frac{\Theta \Lambda}{8\pi G_{N}^{(d)}}
    +\frac{\alpha  r_{h}^{d-1}}{(d-2)\gamma}\,,
  \end{aligned}
\end{equation}

\noindent
where

\begin{subequations}
  \begin{align}
    \Phi
    & \equiv
      P_{k}(r_{h})
      =
\sqrt{\frac{2(d-2)}{(d-3)}} \frac{q}{r_{h}^{d-3}}\,,
    \\
    & \nonumber \\
    \Theta
    & \equiv
      \tfrac{1}{2}\frac{\omega_{(d-2)}r_{h}^{d-1}}{d-1}\,,
  \end{align}
\end{subequations}

\noindent
are the electric potential of the black-hole horizon and the thermodynamic
potential conjugate to the cosmological constant, respectively.

Equating $\mathcal{K}(S_{\infty}^{d-2})$ and $\mathcal{K}(S_{r_{h}}^{d-2})$ we
arrive at the Smarr formula

\begin{equation}
  \label{eq:SmarrCRNT}
  \frac{(d-3)M}{(d-2)}
  =
  \frac{ \kappa A}{8\pi G_{N}^{(d)}}
    +\frac{(d-3)}{(d-2)}\Phi Q
    -2\frac{\Theta \Lambda}{8\pi G_{N}^{(d)}}\,.
\end{equation}

Observe that correct factors in the $\Phi Q$ term are obtained only when the
contribution from the $F^{2}$ term to $\omega^{\alpha\mu}$ and contribution
from the $F^{\alpha\mu}P_{k}$ are taken into account.

\subsection{$f(R)$ gravity}
\label{sec-f(R)gravity}

$f(R)$ gravities,\footnote{For a review with many references, see
  Ref.~\cite{Sotiriou:2008rp}.} although equivalent to a Brans-Dicke theory
with scalar potential, can also be used to illustrate the use of the Komar
integral Eq.~(\ref{eq:Komarintegralgeneral}). It will also help us to make an
interesting point on black-hole thermodynamics.

These theories are defined by a Lagrangian density of the form

\begin{equation}
\mathcal{L} = \frac{\sqrt{|g|}}{16\pi G_{N}^{(d)}}f(R)\,,  
\end{equation}

\noindent
where $f(R)$ is a function of the Ricci scalar. We just need to compute

\begin{subequations}
  \begin{align}
  P^{\alpha\mu\beta\nu}
  & =
    \frac{\sqrt{|g|}}{16\pi G_{N}^{(d)}}g^{\alpha\mu\, \beta\nu}f'\,,
    \\
    & \nonumber \\
  \nabla_{\rho}P^{\alpha\mu\beta\nu}
  & =
    \frac{\sqrt{|g|}}{16\pi G_{N}^{(d)}}g^{\alpha\mu\, \beta\nu}\nabla_{\rho}f'\,,
  \end{align}
\end{subequations}

\noindent
where $f'$ is the derivative of the function $f$ with respect to its argument.

Then, Eq.~(\ref{eq:EinsteinEquations}) gives the Einstein equations

\begin{equation}
  \label{eq:EEf(R)}
\frac{16\pi G_{N}^{(d)}}{\sqrt{|g|}}    \frac{\delta S}{\delta g_{\mu\nu}}
=
\tfrac{1}{2}g^{\mu\nu}f-f' R^{\mu\nu} +(g^{\mu\nu}\nabla^{2}-\nabla^{\mu}\nabla^{\nu})f'\,,
\end{equation}

\noindent
and Eq.~(\ref{eq:Komarintegralgeneral}) gives the Komar integral

\begin{equation}
  \label{eq:Komarintegralf(R)}
    \mathcal{K}(\Sigma^{d-2})
    =
    \frac{(-1)^{d-1}}{16\pi G_{N}^{(d)}}
    \int_{\Sigma^{d-2}}d^{d-2}\Sigma_{\alpha\mu}
    \left\{f'\nabla^{\alpha}k^{\mu}
      -2\nabla^{\alpha}f'k^{\mu}
      -\omega^{\alpha\mu}\right\}\,,
\end{equation}

\noindent
where

\begin{equation}
  \nabla_{\alpha}\omega^{k\,\alpha\mu}
  =
  \frac{16\pi G_{N}{}^{(d)}}{\sqrt{|g|}}\mathcal{L}\,,
\end{equation}

\noindent
where the Lagrangian density has to be evaluated on-shell. The trace of the
equation of motion (\ref{eq:EEf(R)}) gives the relation

\begin{equation}
  \label{eq:traceEEf(R)}
  \frac{d}{2}f-f' R +(d-2)\nabla^{2}f'
  =
  0\,,
\end{equation}

\noindent
which, as in the Lovelock case, can be used to eliminate terms in
$\mathcal{L}$.

Let us consider a simple model, described by a second degree polynomial:

\begin{equation}
  f(R)
  =
  -(d-2)\Lambda +R + \alpha_{2} R^{2}\,,
\end{equation}

\noindent
where the zeroth and linear (Einstein-Hilbert) terms have the standard normalization.

The cosmological Schwarzschild-Tangherlini metrics \cite{Tangherlini:1963bw}
given in Eqs.~(\ref{eq:CRNTsolution}) setting $q=0$, are known to be solutions
of these theories.  The constants $m$ and $\lambda$ are, essentially, the mass
and the cosmological constant in absence of the $R^{2}$ term, according to
Eqs.~(\ref{eq:CRNTphysicalparameters}). In this theory, though, there is a
more complicated relation between $m$ and $\lambda$ and $M$ and $\Lambda$.

\begin{equation}
R_{\mu\nu}= \lambda g_{\mu\nu}\,,
\end{equation}

\noindent
where $\lambda$ is the effective cosmological constant, to be distinguished
from the cosmological constant $\Lambda$ in the action.  Substituting this
condition in Eq.~(\ref{eq:traceEEf(R)}) we find (only for $d\neq 4$ and
$\alpha_{2}\neq 0$)

\begin{equation}
  \lambda
  =
  -\frac{1}{\beta}\left\{1 \pm\sqrt{1 +2\Lambda\beta}\right\}\,,
  \,\,\,\,\,\,
  \text{where}
  \,\,\,\,\,\,
  \beta \equiv 2\alpha_{2} d(d-4)/(d-2)\,.
\end{equation}

For this theory and these solutions, the Komar integral
Eq.~(\ref{eq:Komarintegralf(R)}) over a $S^{d-2}$ of radius $r$

\begin{equation}
    \mathcal{K}(S_{r}^{d-2})
    =
    \frac{\omega_{(d-2)}}{8\pi G_{N}^{(d)}} 
    (1+2\alpha_{2}d\lambda)\left\{\frac{r^{d-2}}{2}W'
      +\frac{\lambda r^{d-1}}{(d-1)}+\alpha r_{h}^{d-1}\right\}\,.
\end{equation}

The terms that diverge at infinity cancel identically and we get

\begin{equation}
    \mathcal{K}(S_{\infty}^{d-2})
    =
     \frac{(d-3)}{(d-2)}M
    +\frac{\alpha r_{h}^{d-1}}{(d-2)\gamma}(1+2\alpha_{2}d\lambda)\,,
\end{equation}

\noindent
where, now

\begin{equation}
  \label{eq:massf(R)}
m = \gamma M/(1+2\alpha_{2}d\lambda)\,.
\end{equation}

Over the horizon

\begin{equation}
    \mathcal{K}(S_{r_{h}}^{d-2})
    =
    ST
    +\frac{\omega_{(d-2)}r_{h}^{d-1}}{8\pi G_{N}^{(d)}(d-1)} 
    (1+2\alpha_{2}d\lambda)\lambda
    +\frac{\alpha r_{h}^{d-1}}{(d-2)\gamma}(1+2\alpha_{2}d\lambda)
\end{equation}

\noindent
where $T=\frac{\kappa}{2\pi}$ and $S$, Wald's entropy, is given by 

\begin{equation}
  \label{eq:entropyf(R)}
S =  (1+2\alpha_{2}d\lambda)\frac{\omega_{(d-2)} r_{h}^{d-2}}{4 G_{N}^{(d)}}\,,
\end{equation}

\noindent
and we finally arrive at

\begin{equation}
  \frac{(d-3)}{(d-2)}M
  =
    ST
    +\frac{\omega_{(d-2)}r_{h}^{d-1}}{8\pi G_{N}^{(d)}(d-1)} 
    (1+2\alpha_{2}d\lambda)\lambda\,. 
\end{equation}

The second term in the right-hand side can be interpreted in the same spirit
as in the Lovelock case, in terms of two contributions associated to the
dimensionful parameters that appear in the action, $\Lambda$ and $\alpha_{2}$,
which will be identified with thermodynamical variables (``pressures'') and
their conjugate potentials (``volumes'') $\Theta_{\Lambda}$ and
$\Theta_{\alpha_{2}}$

\begin{equation}
  \label{eq:BD}
  \frac{(d-3)M}{(d-2)}
  =
  ST
  -2\frac{\Theta_{\Lambda} \Lambda (1+2\alpha_{2}d\lambda)}{8\pi G_{N}^{(d)}}
  +2\frac{\Theta_{\alpha_{2}} \alpha_{2} (1+2\alpha_{2}d\lambda)}{8\pi G_{N}^{(d)}}\,,
\end{equation}

\noindent
with $\Theta_{\Lambda}$ and $\Theta_{\alpha_{2}}$ given by

\begin{subequations}
  \begin{align}
    \Theta_{\Lambda}
    & \equiv
      \frac{\Theta}{1+2\alpha_{2}d\lambda}\,,
    \\
    & \nonumber \\
    \Theta_{\alpha_{2}}
    & \equiv
      - \frac{d^{2}\lambda^{2}\Theta}{(d-2)(1+2\alpha_{2}d\lambda)}\,.
  \end{align}
\end{subequations}

These ``volumes'' are the contributions of the cosmological and $R^{2}$ terms
to the Komar integral on the horizon.

It is interesting to recover the same results in the equivalent Brans-Dicke
theory, whose Lagrangian density has the form

\begin{equation}
  \mathcal{L}
  =
  \frac{\sqrt{|g|}}{16\pi G_{N}^{(d)}}\left[\phi R - V(\phi)\right]\,,    
\end{equation}

\noindent
with

\begin{subequations}
  \begin{align}
    \phi
    & =
      f'(R) = 1+2\alpha_{2}R\,,
    \\
    & \nonumber \\
    V(\phi)
    & =
      Rf'(R) -f(R) = (d-2)\Lambda +\frac{1}{4\alpha_{2}}(\phi-1)^{2}\,.
  \end{align}
\end{subequations}

It is convenient to rescale the metric to the Einstein frame. If we want the
rescaling to relate asymptotically-flat or aDS metrics with the same
normalization, the rescaling has to be performed with $\phi/\phi_{\infty}$,
where $\phi_{\infty}$ is the asymptotic value of the scalar
($1+2\alpha_{2}d\lambda$ in our case). First, we rewrite the Lagrangian as

\begin{equation}
  \mathcal{L}
  =
  \frac{\sqrt{|g|}\phi_{\infty}}{16\pi G_{N}^{(d)}}
  \left[(\phi/\phi_{\infty}) R - V(\phi)/\phi_{\infty}\right]\,,    
\end{equation}

\noindent
and replace the metric $g_{\mu\nu}$ by
$(\phi/\phi_{\infty})^{-2/(d-2)}g_{\mu\nu}$, which leads to

\begin{equation}
  \mathcal{L}
  =
  \frac{\sqrt{|g|}\phi_{\infty}}{16\pi G_{N}^{(d)}}
  \left[R +\frac{(d-2)}{(d-2)}(\partial\ln{(\phi/\phi_{\infty})})^{2}
    -2\frac{(d-2)}{(d-2)}\nabla^{2}\ln{(\phi/\phi_{\infty})}
    - \mathcal{V}(\phi)\right]\,,    
\end{equation}

\noindent
where

\begin{equation}
  \mathcal{V}(\phi)
  \equiv
   \phi_{\infty}^{2/(d-2)}\left[(d-2)\Lambda\phi^{-d/(d-2)}
  +\frac{1}{4\alpha_{2}}\phi^{-d/(d-2)}(\phi-1)^{2}\right]\,,
\end{equation}

\noindent
is the new scalar potential which, not surprisingly, is extremized by
$\phi=1+2\alpha_{2}d\lambda$. This is the value of $\phi_{\infty}$ and we see
that the effective cosmological constant is
$\lambda$,\footnote{$\mathcal{V}(\phi_{\infty})=(d-2)\lambda$ in our
  conventions.} and that the effective Newton constant in this theory is just
$G_{N}^{(d)}/\phi_{\infty}$. The mass and the entropy, measured in this
theory, take, then, the same value as the mass and entropy measured in the
higher-order $f(R)$ theory, Eqs.~(\ref{eq:massf(R)}) and
(\ref{eq:entropyf(R)}).

We can use our previous results to find the Smarr formula through the Komar
integral for this theory. The on-shell Lagrangian is proportional to the
scalar potential and we could express it as in Eq.~(\ref{eq:SmarrCRNT}) with
$\Lambda$ replaced by the effective cosmological constant $\lambda$. However,
it is more natural to express it in terms of the original dimensionful
constants that define the theory, namely $\Lambda$ and $\alpha_{2}$, as in 
Eq.~(\ref{eq:BD}).

In the Brans-Dicke form of the theory, the ``volumes'' can be computed as
surface or volume integrals associated to the two terms of the potential:

\begin{subequations}
  \begin{align}
    \Theta_{\Lambda}
    & =
      \frac{1}{(d-2)}\int_{\mathcal{B}^{d-2}_{r_{h}}}d^{d-1}\Sigma_{\mu}k^{\mu}
      \frac{\partial \mathcal{V}}{\partial \Lambda}\,,
    \\
    & \nonumber \\
    \Theta_{\alpha_{2}}
    & =
      \frac{1}{(d-2)}\int_{\mathcal{B}^{d-2}_{r_{h}}}d^{d-1}\Sigma_{\mu}k^{\mu}
       \frac{\partial \mathcal{V}}{\partial \alpha_{2}}\,,
  \end{align}
\end{subequations}

\noindent
where $\mathcal{B}^{d-2}_{r_{h}}$ is the ball of radius $r_{h}$ whose boundary
is the spatial section of the horizon we have considered.

This suggests that for potentials depending on parameters $\alpha_{k}$ of
dimensions $[L^{k}]$ the Smarr formula should take the general form

\begin{equation}
  \label{eq:SmarrBDgeneral}
  \frac{(d-3)M}{(d-2)}
  =
  ST
  +\sum_{k}k\frac{\Theta_{\alpha_{k}} \alpha_{k} \phi_{\infty}}{8\pi G_{N}^{(d)}}\,,
\end{equation}

\noindent
with 

\begin{equation}
    \Theta_{\alpha_{k}}
 =
      \frac{1}{(d-2)}\int_{\mathcal{B}^{d-2}_{r_{h}}}d^{d-1}\Sigma_{\mu}k^{\mu}
       \frac{\partial \mathcal{V}}{\partial \alpha_{k}}\,.
\end{equation}

This point of view should be contrasted with ethe one in
Ref.~\cite{Kastor:2018cqc} in which all these terms are, effectively, combined
in one.

\section{Discussion}
\label{sec-discussion}

In this paper we have argued that the dimensionful constants that define
scalar potentials should be treated as thermodynamical variables, by analogy
with the treatment of the dimensional parameters of Lovelock theories in
Ref.~\cite{Kastor:2010gq} or of the parameter of the Born-Infeld theory in
Ref.~\cite{Gunasekaran:2012dq}. Although we have not studied directly the
first law of black-hole mechanics, it is clear that the one could proceed, for
instance,as in Ref.~\cite{Urano:2009xn}, including variations of those
parameters, to derive a first law that includes them as thermodynamical
parameters using Wald's formalism. A less \textit{ad hoc} procedure is,
nevertheless, quite desirable and work in this direction is well under way
\cite{kn:EMMOP}.

The most interesting example of theories with scalar fields and scalar
potentials is provided by gauged supergravities.\footnote{See
  Ref.~\cite{Trigiante:2016mnt} for a quite complete review and
  Ref.~\cite{Ortin:2015hya} for a more pedagogical introduction.} Their
potentials depend on coupling constants that ``deform'' the original theory,
such as gauge coupling constants and St\"uckelberg mass parameters. The
implication is that those parameters, many of them codified in the so-called
\textit{embedding tensor} should be considered thermodynamical
variables. These parameters can, in their turn, be related to potentials that
couple to branes \cite{Bergshoeff:2009ph,Hartong:2009vc} which raises the
possibility of intriguing connections between those brase and the
thermodynamical variables which deserve to be studied \cite{kn:EMMOP}.

Since, in Section~\ref{sec-CRNT} we have considered charged black holes in
general $d$ dimensions, we have not considered the magnetically charged ones
that can exist in $d=4$. Those solutions can easily be obtained using
electric-magnetic duality. In particular we should replace $q^{2}$ by
$q^{2}+p^{2}$, where $p$ is (up to constants) the magnetic charge in the
metric and, correspondingly, $\Phi Q$ by $\Phi Q+ \Xi P$, where $\Xi$ is the
magnetic potential on the horizon and $P$ the magnetic charge, in the Smarr
formula. However, had we considered those solutions, and had we used the Komar
formula Eq.~(\ref{eq:Komarintegralgeneral}) we would have arrived to a
different (wrong) Smarr formula. This is due to the fact that Wald's formalism
can only account for variations of electric charges\footnote{Those which would
  be naturally carried by objects coupling to the potentials appearing in the
  action. The Maxwell field, for instance, couples to electrically-charged
  point particles only. Magentic monopoles couple to the dual of the Maxwell
  field.} in the first law, a problem we have pointed out in
Ref.~\cite{Elgood:2020mdx} and which shows in the expressions for the first
law obtained in Refs.~\cite{Elgood:2020svt,Elgood:2020mdx,Elgood:2020nls}.

Although they do not contribute to the Smarr fomulae,\footnote{This can be
  seen using the scaling argument reviewed, for instance, in
  Ref.~\cite{Kastor:2010gq}.} scalars do contribute to the first law
\cite{Gibbons:1996af,Astefanesei:2018vga} and Wald's formalism cannot account
for their contributions, either. A common solution to the two problems that we
have mentioned (and, perhaps, to properly include the variations of the
dimensionful constants) would be to use ``democratic formulations'' of the
theories under consideration, which include all the original (``fundamental''
or ``electric'') fields of the theory (including deformation constants)
together with their (``magnetic'') duals, as in
Ref.~\cite{Bergshoeff:2001pv,Bergshoeff:2009ph}. Work in this direction is
also in progress \cite{kn:EMMO}.

\section*{Acknowledgments}

The author would like to thank R.B.~mann for the course on black-hole
chemistry given at the IFT and P.~Bueno, P.~Cano, P.~Meessen and
D.~Pere\~n\'{\i}guez for useful conversations.  This work has been supported
in part by the MCIU, AEI, FEDER (UE) grant PGC2018-095205-B-I00 and by the
Spanish Research Agency (Agencia Estatal de Investigaci\'on) through the grant
IFT Centro de Excelencia Severo Ochoa SEV-2016-0597.  TO wishes to thank
M.M.~Fern\'andez for her permanent support.

\appendix

\section{Determination of $J^{\alpha\mu}$}
\label{app-calculations}

Let us consider the first term in Eq.~(\ref{eq:Jmu}). Expanding the
symmetrizer and using the Ricci identity

\begin{equation}
  [\nabla_{\mu},\nabla_{\nu}]\xi_{\rho}
  =
  -R_{\mu\nu\rho\sigma}\xi^{\sigma}\,,
\end{equation}

\noindent
we get

\begin{equation}
  \begin{aligned}
    P^{\mu\alpha\beta\gamma}\nabla_{\beta}\nabla_{(\alpha}\xi_{\gamma)}
    & =
    \tfrac{1}{2}P^{\mu\alpha\beta\gamma}\nabla_{\beta}\nabla_{\alpha}\xi_{\gamma}
    +\tfrac{1}{2}P^{\mu\alpha\beta\gamma}\nabla_{\beta}\nabla_{\gamma}\xi_{\alpha}
    \\
    & \\
    & =
    \tfrac{1}{2}P^{\mu\alpha\beta\gamma}\left(\nabla_{\alpha}\nabla_{\beta}\xi_{\gamma}
      -R_{\beta\alpha\gamma\nu}\xi^{\nu}\right)
    +\tfrac{1}{2}P^{\mu\alpha\beta\gamma}
    \left(-\tfrac{1}{2}R_{\beta\gamma\alpha\nu}\xi^{\nu}\right)
    \\
    & \\
    & =
    \tfrac{1}{2}P^{\mu\alpha\beta\gamma}\nabla_{\alpha}\nabla_{\beta}\xi_{\gamma}
    +\tfrac{1}{2}P^{\mu\alpha\beta\sigma}R_{\nu\alpha\beta\sigma}\,,
  \end{aligned}
\end{equation}

\noindent
where we have used the Bianchi identity $R_{[\mu\nu\rho]\sigma}=0$ in the last
step. Integrating by parts the first term twice, we get

\begin{equation}
  \begin{aligned}
    P^{\mu\alpha\beta\gamma}\nabla_{\beta}\nabla_{(\alpha}\xi_{\gamma)}
    & =
    \nabla_{\alpha}\left(\tfrac{1}{2}P^{\mu\alpha\beta\gamma}\nabla_{\beta}\xi_{\gamma}\right)
    -\tfrac{1}{2}\nabla_{\alpha}P^{\mu\alpha\beta\gamma}\nabla_{\beta}\xi_{\gamma}
    +\tfrac{1}{2}P^{\mu\alpha\beta\sigma}R_{\nu\alpha\beta\sigma}
    \\
    & \\
    & =
    \nabla_{\alpha}\left(\tfrac{1}{2}P^{\mu\alpha\beta\gamma}\nabla_{\beta}\xi_{\gamma}\right)
    +\nabla_{\beta}\left(-\tfrac{1}{2}\nabla_{\alpha}P^{\mu\alpha\beta\gamma}\xi_{\gamma}\right)
    \\
    & \\
    & \hspace{.5cm}
    +\tfrac{1}{2}\nabla_{\beta}\nabla_{\alpha}P^{\mu\alpha\beta\gamma}\xi_{\gamma}
    +\tfrac{1}{2}P^{\mu\alpha\beta\sigma}R_{\nu\alpha\beta\sigma}
    \\
    & \\
    & =
    \nabla_{\alpha}\left(\tfrac{1}{2}P^{\mu\alpha\beta\gamma}\nabla_{\beta}\xi_{\gamma}\right)
    +\nabla_{\beta}\left(-\tfrac{1}{2}\nabla_{\alpha}P^{\mu\alpha\beta\gamma}\xi_{\gamma}\right)
    \\
    & \\
    & \hspace{.5cm}
    +\nabla_{(\beta}\nabla_{\alpha)} P^{\mu\alpha\beta\gamma}\xi_{\gamma}
    -\tfrac{1}{2}\nabla_{\alpha}\nabla_{\beta}P^{\mu\alpha\beta\gamma}\xi_{\gamma}
    +\tfrac{1}{2}P^{\mu\alpha\beta\sigma}R_{\nu\alpha\beta\sigma}
    \\
    & \\
    & =
    \nabla_{\alpha}\left(\tfrac{1}{2}P^{\mu\alpha\beta\gamma}\nabla_{\beta}\xi_{\gamma}-\tfrac{1}{2}\nabla_{\beta}P^{\mu\alpha\beta\gamma}\xi_{\gamma}\right)
    +\nabla_{\beta}\left(-\tfrac{1}{2}\nabla_{\alpha}P^{\mu\alpha\beta\gamma}\xi_{\gamma}\right)
    \\
    & \\
    & \hspace{.5cm}
    +\nabla_{(\beta}\nabla_{\alpha)} P^{\mu\alpha\beta\gamma}\xi_{\gamma}
    +\tfrac{1}{2}\nabla_{\beta}P^{\mu\alpha\beta\gamma}\nabla_{\alpha}\xi_{\gamma}
    +\tfrac{1}{2}P^{\mu\alpha\beta\sigma}R_{\nu\alpha\beta\sigma}\,,
  \end{aligned}
\end{equation}

\noindent
where we have performed yet another integration by parts in the last step.

Now we have to use the identity Eq.~(\ref{eq:BianchiP}) in the second term of
the first line

\begin{equation}
  \begin{aligned}
    P^{\mu\alpha\beta\gamma}\nabla_{\beta}\nabla_{(\alpha}\xi_{\gamma)}
    & =
    \nabla_{\alpha}\left(\tfrac{1}{2}P^{\mu\alpha\beta\gamma}\nabla_{\beta}\xi_{\gamma}-\tfrac{1}{2}\nabla_{\beta}P^{\mu\alpha\beta\gamma}\xi_{\gamma}\right)
    \\
    & \\
    & \hspace{.5cm}
    +\nabla_{\beta}\left[\tfrac{1}{2}\nabla_{\alpha}
      \left(P^{\alpha\beta\mu\gamma}+P^{\beta\mu\alpha\gamma}\right)\xi_{\gamma}\right]
    \\
    & \\
    & \hspace{.5cm}
    -\nabla_{(\alpha}\nabla_{\beta)} P^{\mu\alpha\beta\gamma}\xi_{\gamma}
    +\tfrac{1}{2}\nabla_{\beta}P^{\mu\alpha\beta\gamma}\nabla_{\alpha}\xi_{\gamma}
    +\tfrac{1}{2}P^{\mu\alpha\beta\sigma}R_{\nu\alpha\beta\sigma}
    \\
    & \\
    & =
    \nabla_{\alpha}\left(\tfrac{1}{2}P^{\mu\alpha\beta\gamma}\nabla_{\beta}\xi_{\gamma}-\tfrac{1}{2}\nabla_{\beta}P^{\mu\alpha\beta\gamma}\xi_{\gamma}\right)
    \\
    & \\
    & \hspace{.5cm}
    -\nabla_{\beta}\left(\tfrac{1}{2}\nabla_{\alpha}P^{\beta\alpha\mu\gamma}\xi_{\gamma}\right)
        -\nabla_{\alpha}\left(\tfrac{1}{2}\nabla_{\beta}P^{\mu\alpha\beta\gamma}\xi_{\gamma}\right)
    \\
    & \\
    & \hspace{.5cm}
    +\tfrac{1}{2}\nabla_{\beta}P^{\mu\alpha\beta\gamma}\nabla_{\alpha}\xi_{\gamma}
        -\nabla_{(\alpha}\nabla_{\beta)} P^{\mu\alpha\beta\gamma}\xi_{\gamma}
    +\tfrac{1}{2}P^{\mu\alpha\beta\sigma}R_{\nu\alpha\beta\sigma}
    \\
    & \\
    & =
    \nabla_{\alpha}\left(\tfrac{1}{2}P^{\mu\alpha\beta\gamma}\nabla_{\beta}\xi_{\gamma}
      -\nabla_{\beta}P^{\mu\alpha\beta\gamma}\xi_{\gamma}\right)
    \\
    & \\
    & \hspace{.5cm}
    -\tfrac{1}{2}\nabla_{\beta}\nabla_{\alpha}P^{\beta\alpha\mu\gamma}\xi_{\gamma}
    \\
    & \\
    & \hspace{.5cm}
      +\nabla_{\beta}P^{\beta\alpha\mu\gamma}\nabla_{(\alpha}\xi_{\gamma)}
        -\nabla_{(\alpha}\nabla_{\beta)} P^{\mu\alpha\beta\gamma}\xi_{\gamma}
    +\tfrac{1}{2}P^{\mu\alpha\beta\sigma}R_{\nu\alpha\beta\sigma}\,.
  \end{aligned}
\end{equation}

Now, because of the antisymmetry of $P^{\mu\nu\rho\sigma}$ in the first two
indices, we can use the Ricci identity to show that the term in the second line
vanishes identically:

\begin{equation}
  \begin{aligned}
    \nabla_{\alpha}\nabla_{\beta}P^{\alpha\beta\mu\gamma}
    & =
    \tfrac{1}{2} \left\{
      R_{\alpha\beta\delta}{}^{\alpha}P^{\delta\beta\mu\gamma}
      +R_{\alpha\beta\delta}{}^{\beta}P^{\alpha\delta\mu\gamma}
      +R_{\alpha\beta\delta}{}^{\mu}P^{\alpha\beta\delta\gamma}
      +R_{\alpha\beta\delta}{}^{\gamma}P^{\alpha\beta\mu\delta} \right\}
    \\
    & \\
    & =
    \tfrac{1}{2} \left\{
      -R_{\beta\delta}P^{\delta\beta\mu\gamma}
      +R_{\alpha\delta}P^{\alpha\delta\mu\gamma}
      +R_{\alpha\beta\delta}{}^{\mu}P^{\alpha\beta\delta\gamma}
      +R_{\alpha\beta\delta}{}^{\gamma}P^{\alpha\beta\mu\delta} \right\}
    \\
    & \\
    & =
      R_{\alpha\beta\delta}{}^{(\mu|}P^{\alpha\beta\delta|\gamma)}
    \\
    & \\
    & =
    0\,,
  \end{aligned}
\end{equation}

\noindent
where we have used Eq.~(\ref{eq:PRproperty}) in the last step.

Flipping the indices $\alpha\mu$, we arrive to the identity

\begin{equation}
  \begin{aligned}
    \nabla_{\alpha}\left(\nabla_{\beta}P^{\alpha\mu\beta\gamma}\xi_{\gamma}
      -\tfrac{1}{2}P^{\alpha\mu\beta\gamma}\nabla_{\beta}\xi_{\gamma}
    \right)
    & =
    P^{\mu\alpha\beta\gamma}\nabla_{\beta}\nabla_{(\alpha}\xi_{\gamma)}
    -\nabla_{\beta}P^{\beta\alpha\mu\gamma}\nabla_{(\alpha}\xi_{\gamma)}
    \\
    & \\
    & \hspace{.5cm}
    -\tfrac{1}{2}P^{\mu\alpha\beta\sigma}R_{\nu\alpha\beta\sigma}
   +\nabla_{(\alpha}\nabla_{\beta)} P^{\mu\alpha\beta\gamma}\xi_{\gamma}\,,
      \end{aligned}
\end{equation}

\noindent
from which Eq.~(\ref{eq:Jalphamu}) easily follows.


\end{document}